\documentclass[conference]{IEEEtran}
\IEEEoverridecommandlockouts
\usepackage{cite}
\usepackage{amsmath,amssymb,amsfonts}
\usepackage{algorithmic}
\usepackage{graphicx}
\usepackage{textcomp}
\usepackage{xcolor}

\usepackage{url}

\def\BibTeX{{\rm B\kern-.05em{\sc i\kern-.025em b}\kern-.08em
    T\kern-.1667em\lower.7ex\hbox{E}\kern-.125emX}}
\begin{document}

\title{Understanding Toxic Interaction Across User and Video Clusters in Social Video Platforms
\thanks{This work was supported by JST SPRING Grant Number JPMJSP2124 and JST-Mirai Program Grant Number JPMJMI23B1, Japan.}
}

\author{
\IEEEauthorblockN{Qiao Wang\IEEEauthorrefmark{1},
Liang Liu\IEEEauthorrefmark{1},
and Mitsuo Yoshida\IEEEauthorrefmark{2}}
\IEEEauthorblockA{\IEEEauthorrefmark{1}Degree Programs in Systems and Information Engineering, University of Tsukuba, Tsukuba, Japan \\
Email: \{s2430137, s2330142\}@u.tsukuba.ac.jp}
\IEEEauthorblockA{\IEEEauthorrefmark{2}Institute of Business Sciences, University of Tsukuba, Tokyo, Japan \\
Email: mitsuo@gssm.otsuka.tsukuba.ac.jp}
}

\maketitle

\begin{abstract}
Social video platforms shape how people access information, while recommendation systems can narrow exposure and increase the risk of toxic interaction. 
Previous research has often examined text or users in isolation, overlooking the structural context in which such toxic interactions occur. 
Without considering who interacts with whom and around what content, it is difficult to explain why negative expressions cluster within particular communities.
To address this issue, this study focuses on the Chinese social video platform Bilibili, incorporating video-level information as the environment for user expression, modeling users and videos in an interaction matrix.
After normalization and dimensionality reduction, we perform separate clustering on both sides of the video–user interaction matrix with K-means. 
Cluster assignments facilitate comparisons of user behavior, including message length, posting frequency, and source (barrage and comment), as well as textual features such as sentiment and toxicity, and video attributes defined by uploaders. 
Such a clustering approach integrates structural ties with content signals to identify stable groups of videos and users. 
We find clear stratification in interaction style (message length, comment ratio) across user clusters, while sentiment and toxicity differences are weak or inconsistent across video clusters.
Across video clusters, viewing volume exhibits a clear hierarchy, with higher exposure groups concentrating more toxic expressions. 
For such a group, platforms should require timely intervention during periods of rapid growth. 
Across user clusters, comment ratio and message length form distinct hierarchies, and several clusters with longer and comment-oriented messages exhibit lower toxicity.
For such groups, platforms should strengthen mechanisms that sustain rational dialogue and encourage engagement across topics.
\end{abstract}


\section{Introduction}
With the rapid growth of the Internet, social video platforms have become key venues for both information acquisition and social interaction. Their recommendation systems rank and push content according to observable user behaviors such as clicks, viewing duration, and likes~\cite{davidson2010youtube}. 
This mechanism undoubtedly strengthens user engagement and retention, but it also increases the frequency and fragmentation of information exposure. 
Empirical studies show that such systems may foster information cocoons or filter bubbles, in which users are repeatedly exposed to homogeneous content~\cite{hartmann2025systematic}. 
Prolonged exposure to uniform information environments has been linked to the polarization of opinions~\cite{hou2023information}. 
Within polarized groups, individuals are more prone to express intense emotions and toxic speech, creating both psychological strain and risks of social conflict. 
These issues highlight the urgent need to examine how the frequency of toxic interactions relates to both video content and user behavior.

Research on toxic interactions has produced substantial findings in the areas of toxicity detection and user governance. 
One perspective emphasizes textual analysis, detecting and interpreting hateful or toxic speech by modeling linguistic features, contextual cues, and latent emotions~\cite{ali2022hate, mossie2020vulnerable, ghenai2025exploring}. 
Another perspective highlights user-centered perspectives, focusing on individual profiles, behavioral attributes, and situational triggers~\cite{ribeiro2018characterizing, cheng2017anyone, geissler2025analyzing}. 
While both approaches offer valuable insights, they also have limitations. 
Text-based models alone cannot explain why negative interactions concentrate within particular groups or communities. 
User-based models, in turn, have difficulty revealing how recommendation mechanisms amplify emotional polarization and sustain controversy. 
In short, the structural dimension of who interacts with whom and around what content remains insufficiently addressed.

This gap is particularly salient on social video platforms that feature both barrage and comment systems. 
Barrages are short text messages displayed on the video screen in real time, synchronized with playback. 
Their brevity and immediacy foster a sense of collective viewing and can readily trigger emotional contagion and imitative behaviors~\cite{wang2025dynamic}. 
By contrast, comments are asynchronous, typically longer and more coherent, and embedded in relatively stable contexts. 
The coexistence of these two modalities reinforces shared interests, also intensifying disputes around controversial topics. 
Thus, these structural distinctions complicate efforts to understand toxic interactions if analysis relies solely on textual features.

These challenges require an analytical lens that connects textual expression to its structural context. 
On social video platforms, users’ sentiments and behaviors are embedded in interactions surrounding specific content. 
Incorporating this video dimension allows the analysis to capture how sentiment and toxicity arise within shared viewing environments. 
Accordingly, we model videos and users as two sets of nodes in a bipartite network and construct a video–user interaction matrix for separate clustering analysis.
Specifically, the user-by-user and user-by-video matrices are normalized and subjected to dimensionality reduction before implementing K-means clustering. 
The resulting clusters are then examined to compare user behaviors, such as comment length, posting frequency, and comment source, as well as textual features including sentiment and toxicity scores, and video attributes defined by the uploaders.
This clustering approach aims to explain why trending videos attract more negative interactions, as well as the reasons for distinct emotional patterns among different user groups.
The contributions of this paper are reflected in three aspects:
\begin{itemize}
\item Employs a clustering approach based on the user–video interaction matrix, integrating structural and textual information to analyze group-level patterns of toxic behavior across barrage and comment systems.
\item By comparing the clustering outcomes, we reveal differences between video clusters and user clusters, clarifying how emotional expression and toxicity vary across different interaction mechanisms.
\item Based on the behavioral characteristics of different groups, it summarizes differentiated strategies for platform governance and recommendation system design, providing empirical evidence to encourage constructive discussion and limit harmful interactions.
\end{itemize}

\section{Literature Review}
In recent years, research on toxic interaction, especially hate speech in social media, has taken multiple directions. 
Early studies focused mainly on text, relying on keywords, sentiment features, or deep learning classifiers to detect hate speech and offensive expressions~\cite{ali2022hate, mossie2020vulnerable}. 
These approaches achieved notable accuracy in distinguishing harmful texts, but their reliance on textual signals alone makes it difficult to explain why harmful interactions repeatedly emerge in certain topics or groups.

As the field expanded, attention shifted to examining user level attributes and behavioral patterns.
Empirical work has shown that psychological traits, behavioral patterns, and anonymity strongly shape online communication~\cite{noorian2024user, christopherson2007positive}. 
Other studies demonstrate that fake identities and cross-platform migration intensify the spread and persistence of malicious content~\cite{ramalingam2018fake, ali2021understanding}. 
These findings highlight that perspectives based on content are insufficient, as user-level differences also drive the production and recurrence of toxic expressions. 
However, a perspective centred solely on the user tends to overlook the interactional context in which comments are embedded.

A further body of work examines platform mechanisms, showing that interactions across social media often produce echo chamber effects and homogenized opinions~\cite{cinelli2021echo}. 
These outcomes reflect the combined influence of algorithmic design and user choices. 
Although such studies reveal a platform effect at the macro level, they seldom probe the specific interaction structures through which close circles of users and videos emerge. 

Addressing this limitation, this study turns to Chinese social video platforms, where a noteworthy feature is the coexistence of barrages and traditional comments.
Barrage is often described as asynchronous communication experienced synchronously, affording viewers a temporal sense of live co-presence distinct from the delayed, structured exchanges of the traditional comment section~\cite{zhang2020making}.
Their parallel existence creates a distinctive interaction environment, and relying on data from only one system risks overlooking the full scope of user dynamics.

In summary, existing research has offered insights into content identification, user behavior, and platform effects, while gaps remain. 
On the one hand, interactions between users and videos are often simplified, obscuring the structural distribution of toxic speech within specific circles. 
On the other hand, the distinct roles of comments and barrages have not been fully compared, despite their importance on Chinese social video platforms. 

\section{Methodology}
\subsection{Data Description}
This study investigates which types of content are more likely to trigger toxic interactions among which groups, adopting an interaction structure perspective. 
We focus on Bilibili, a representative Chinese social video platform that integrates barrages and comments as two interaction modes.
Barrages are short, real-time text displayed directly on the video screen and synchronized with playback. 
Their brevity and immediacy create a sense of synchronous viewing that can readily foster emotional contagion and imitation. 
Comments, in contrast, are asynchronous, longer, and contextually stable, resembling a post hoc discussion. 
Comparing these two modes provides a clearer view of how user and video interactions unfold.

As an empirical case, we examine a reality show that generated significant controversy on Bilibili. 
Relevant videos were identified using related keywords, and those lacking either barrages or comments were excluded. 
For eligible videos, both barrage and comment data were collected. 
Data preprocessing was then conducted. 
We retained active users who produced more than five entries, encompassing both barrages and comments, as well as videos with high engagement that accumulated over ten entries.
After removing records with missing values, we ultimately obtained 581 videos and 204,756 text samples contributed by 19,184 users. 
The text dataset includes 160,390 barrages and 44,366 comments.

To capture group-level differences, we constructed user behavior metrics and video interaction metrics. 
User behavior metrics include average barrage and comment frequency, mean message length, sentiment and toxicity scores, and preferred video genres. 
Video metrics include likes, favorites, barrages, and comments. 
To measure relative engagement intensity, we normalized these behaviors by the number of views, producing a per-view engagement rate.
This approach is consistent with prior research that uses likes per view to assess YouTube engagement~\cite{park2016data}.
This study extends this logic to multiple behaviors, providing a more comprehensive characterization of video interaction patterns.

\subsection{Toxicity and Sentiment Analysis}
All barrage and comments collected in this study underwent quantitative toxicity and sentiment analysis.

For toxicity analysis, this study used the Google Perspective API\footnote{\url{https://perspectiveapi.com/}}. 
This tool relies on machine learning to detect potentially harmful content. 
It scores text based on overall toxicity, severe toxicity, insults, threats, identity attacks, and profanity in six dimensions. 
The resulting score ranges from 0 to 1, with higher scores indicating a higher likelihood of the text containing the corresponding harmful characteristics. 
Furthermore, to validate the tool's applicability in the Chinese text, we further used ChatGPT-4o\footnote{\url{https://chatgpt.com/}} as an auxiliary evaluation method, conducting a comparative analysis of the same sample size of barrage and comments. 
Across 1,854 samples, Pearson correlation indicated significant consistency between the two methods ($p < 0.001$).
All six dimensions had moderate positive correlations ($0.4 \le r \le 0.7$), with the highest for insult ($r = 0.60$) and toxicity ($r = 0.55$). 
Such results indicate stable agreement between ChatGPT and the Perspective API in detecting toxic content.
Thus, the Perspective API provides a stable and reliable measure of toxicity in this study. 
Although the Perspective API outputs scores for multiple dimensions, considering language adaptability, overall toxicity was selected as the primary metric in the final analysis, with the remaining dimensions used only for exploratory comparisons.

For sentiment analysis, this paper uses HanLP\footnote{\url{https://www.hanlp.com/}}~\cite{He2021}.
This tool is a deep learning model trained using a multi-task learning framework that has demonstrated strong performance in Chinese text processing. 
HanLP assigns a sentiment score between -1 and 1 to each text entry, with higher values indicating more positive sentiment and lower values indicating more negative sentiment.

After the above preprocessing and calculations, all text entries were assigned corresponding toxicity and sentiment scores. 
Since this study analyzed active users, each of whom posted at least five text entries, we first calculated the average toxicity and sentiment scores for each user. 
Subsequently, we grouped users based on the clustering results and aggregated them at the cluster level to obtain the average toxicity and sentiment scores for each cluster.

\subsection{Statistical and Analytical Methods}
To reveal the structural relationship between users and videos, this study constructed a video-by-user matrix and a user-by-video matrix, respectively. 
The former has videos as rows and users as columns, with elements representing the frequency of message entries a user has made on a particular video. 
The latter has users as rows and videos as columns, with elements representing the number of message entries a user has made on different videos. 
Both matrices were normalized using StandardScaler and pre-dimensionality reduced using principal component analysis. 
These two processing steps mitigated the high-dimensional sparsity of the data and improved computational efficiency. 
Based on the dimensionality reduction results, we used t-SNE to obtain two-dimensional embeddings to visualize the underlying distribution structure of users and videos. 
For clustering, we chose the K-means method. 
This clustering algorithm is simple and computationally efficient, effectively identifying dense clusters of samples, and is therefore suitable for this study. 
We determined the number of clusters for K-means by the Elbow method based on inertial inflection points.

After clustering is complete, cluster evaluation and cluster-level feature induction are performed. 
The clustering quality was quantitatively evaluated using the Silhouette coefficient, the Calinski–Harabasz index, and the Davies–Bouldin index. 
After obtaining user and video cluster labels, we aggregate relevant features at the cluster level, primarily into the following four categories:
\begin{itemize}
    \item User behavior: average number of messages, average message length, and message source distribution (barrage or comments);
    \item Text features: average sentiment and toxicity scores;
    \item Video features: average and total views, likes, favorites, barrage, comments, and interaction rate (interactions/views);
    \item Content category: distribution of typenames within different clusters.
\end{itemize}

Finally, we assessed differences among clusters with non-parametric tests. 
For each behavioral and affective variable at the video and user levels, we ran a Kruskal–Wallis test. 
When the result was significant, we applied Dunn’s pairwise comparisons with Benjamini–Hochberg correction.
We report adjusted p-values ($p_{\mathrm{FDR}}$) and the median difference ($\Delta$) between groups. 
Here, ($\Delta$) equals the median of group$_i$ minus the median of group$_j$.  
Meanwhile, to avoid circular inference, variables used for clustering were standardized without toxicity. 
Toxicity did not enter cluster formation. We examined toxicity only in post hoc comparisons.

\section{Results}
\subsection{Overview of Clustering Results}

The overall evaluation metrics for the video and user clustering models are presented in Table \ref{tab:cluster_quality}.
Their detailed implications will be discussed in the following subsections.

\begin{table}[t]
\centering
\caption{Clustering evaluation results.} 
\label{tab:cluster_quality}
\resizebox{\columnwidth}{!}{
\begin{tabular}{|c|c|c|c|c|}
\hline
\textbf{Type} & \textbf{Clusters} & \textbf{Silhouette} & \textbf{Calinski--Harabasz} & \textbf{Davies--Bouldin} \\
\hline
Video & 8 & 0.348 & 692.620 & 0.882 \\
\hline
User & 9 & 0.353 & 16,202.320 & 0.831 \\
\hline
\end{tabular}
}
\begin{flushleft}
{\footnotesize \textit{Note:} All values are computed based on the final cluster solutions.}
\end{flushleft}
\end{table}

\begin{figure}[tp]
\centering
\includegraphics[width=\columnwidth]{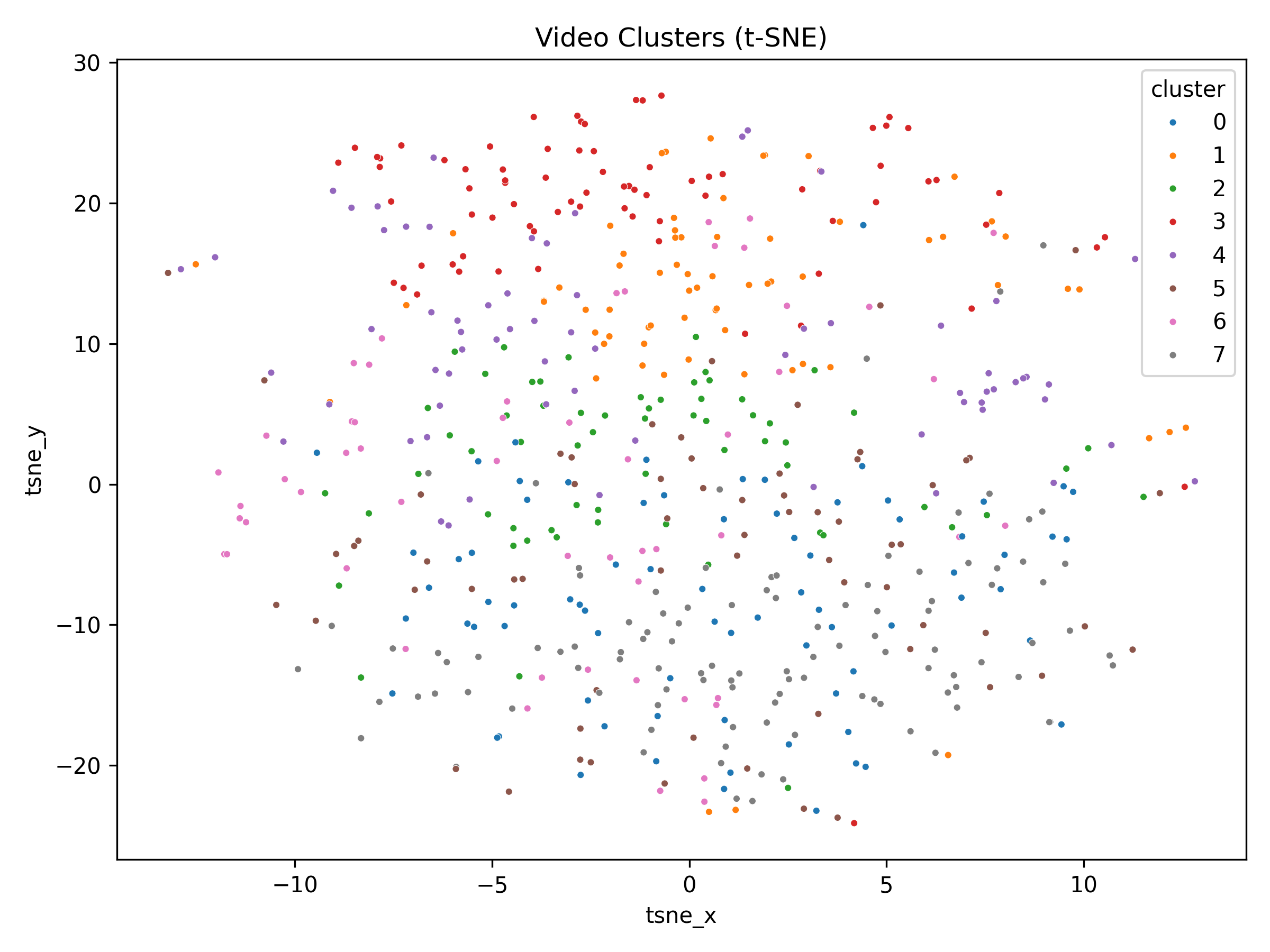}
\caption{t-SNE visualization of video clusters. Colors indicate different cluster IDs in the two-dimensional embedding space.}
\label{fig:tsne_video}
\end{figure}

\begin{figure}[tp]
\centering
\includegraphics[width=\columnwidth]{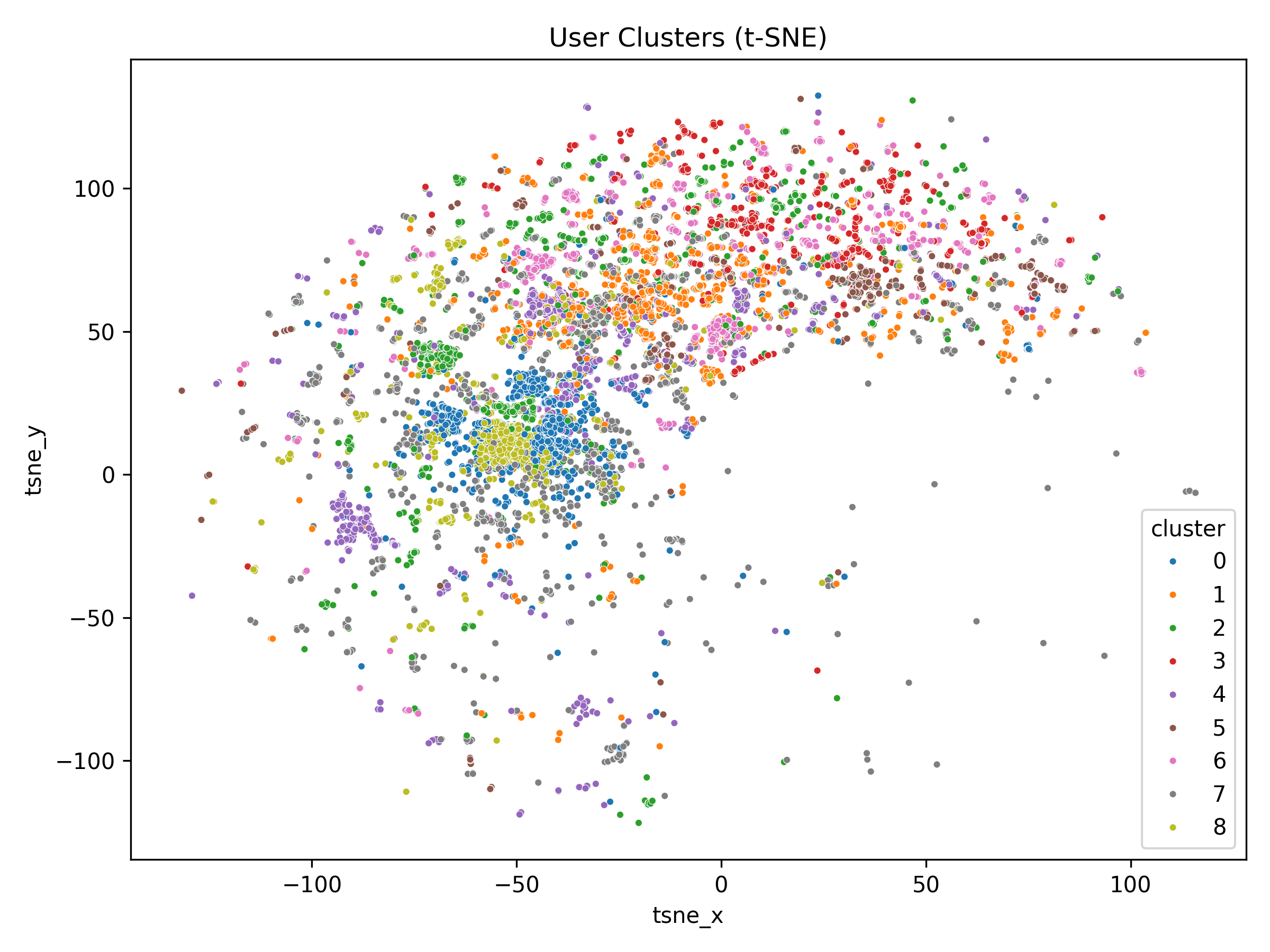}
\caption{t-SNE visualization of user clusters. Colors indicate different cluster IDs in the two-dimensional embedding space.}
\label{fig:tsne_user}
\end{figure}

Meanwhile, to illustrate the clustering structure, we visualized user embeddings using t-SNE.
The visualization reveals clearly separable regions, indicating meaningful differentiation across both video and user clusters.
The video distribution is presented in Fig.~\ref{fig:tsne_video}, and the user distribution in Fig.~\ref{fig:tsne_user}.

\subsection{Video Clustering Results}

\begin{table*}[t]
\caption{Video Cluster Characteristics.}
\begin{center}
\begin{tabular}{|c|c|c|c|c|c|c|c|c|c|c|}
\hline
\textbf{Cluster} & \textbf{Videos} & \textbf{Views} & \textbf{Like} & \textbf{Favorite} & \textbf{Barrage} & \textbf{Comment} & \textbf{Message Length} & \textbf{Sentiment}& \textbf{Toxicity} \\
\hline
0 & 76  & 13,781  & 12.81 & 2.37 & 0.48 & 0.39  & 20.83 & 0.30 & 0.13 \\
1 & 68  & 72,903  & 10.64 & 2.02 & \textbf{1.19} & 0.37  & 23.84 & 0.25 & 0.15 \\
2 & 63  & 19,877  & 7.00  & 1.22 & 0.58 & 0.33  & 26.58 & 0.28 & 0.14 \\
3 & 78  & \textbf{321,372} & 10.83 & 1.53 & 0.45 & 0.36  & 26.58 & 0.23 & \textbf{0.16} \\
4 & 68  & 76,141  & 11.18 & 2.02 & 0.82 & 0.39  & 23.26 & 0.23 & 0.15 \\
5 & 65  & 27,213  & 20.47 & 1.87 & 0.38 & 0.37  & 19.14 & 0.23 & 0.13 \\
6 & 54  & 40,868  & 2.08  & 0.67 & 0.59 & 0.30  & \textbf{42.12} & 0.24 & 0.15 \\
7 & \textbf{109} & 14,237  & \textbf{25.40} & \textbf{3.42} & 0.58 & \textbf{0.41}  & 18.02 & \textbf{0.32} & 0.12 \\
\hline
\end{tabular}
\label{tab:video_cluster_profiles}
\end{center}
\begin{flushleft}
{\footnotesize \textit{Note:} Like, Favorite, Barrage, and Comment are interaction rates, calculated as the ratio of each interaction type to total Views.
Message Length is the average number of characters per message, and Sentiment and Toxicity are the mean scores of all video messages.
The values highlighted in bold represent the highest values across each evaluation index.}
\end{flushleft}
\end{table*}

\begin{table}[t]
\caption{Kruskal–Wallis Test Across Video Clusters.}
\begin{center}
\begin{tabular}{|c|c|c|c|}
\hline
\textbf{Variable} & \textbf{H} & \textbf{$p_{\mathrm{overall}}$} & \textbf{$\varepsilon^2$} \\
\hline
Views         & 358.546 & $<10^{-70}$ & 0.612 \\
Like rate       & 84.193  & $1.92\times 10^{-15}$ & 0.134 \\
Favorite rate   & 30.143  & $8.94\times 10^{-5}$  & 0.040 \\
Barrage rate    & 87.629  & $3.79\times 10^{-16}$ & 0.140 \\
Comment rate    & 31.447  & $5.14\times 10^{-5}$  & 0.043 \\
Message length  & 72.297  & $5.07\times 10^{-13}$ & 0.114 \\
Sentiment       & 9.480   & 0.220                 & 0.004 \\
Toxicity        & 11.281  & 0.127                 & 0.007 \\
\hline
\end{tabular}
\label{tab:overall_video}
\end{center}
\begin{flushleft}
{\footnotesize 
\textit{Note:} Kruskal-Wallis Tests were conducted at the video level. Except for sentiment and toxicity, the remaining attributes are significant in each video cluster. Moreover, the views showed the largest effect sizes in video clustering. }
\end{flushleft}
\end{table}

Based on the video-by-user interaction matrix, the videos are divided into eight clusters. 
The clustering evaluation metrics are summarized in Table~\ref{tab:cluster_quality}. 
The silhouette coefficient was 0.348, the Calinski–Harabasz index was 692.62, and the Davies–Bouldin index was 0.882. 
These results indicate differences among the eight clusters in terms of interaction scale and content categories.
In addition, Table~\ref{tab:video_cluster_profiles} reports user interaction characteristics for each video cluster, comprising interaction behavior, message length, sentiment, and toxicity. 
In this table, Like, Favorite, Barrage, and Comment are interaction rates, calculated as the ratio of each interaction type to total views.
Message Length is the average number of characters per message, and Sentiment and Toxicity are the mean scores of all video messages.
Meanwhile, this paper also uses the Kruskal–Wallis test to evaluate each attribute above. 
The test results show that, except for sentiment and toxicity, the remaining attributes are significant in each video cluster, as shown in Table~\ref{tab:overall_video}.
Views showed the largest omnibus effect in video clustering. Dunn's tests with BH correction revealed clear stratification for views, barrage rate, and like rate.
\begin{itemize}
  \item According to views, clusters followed the order $3 \gg 1 \approx 4 > 6 > 2 \approx 5 > 0 > 7$. For example, cluster~3 far exceeded cluster~7/0/2($\Delta$ = 2000--2700, all $p_{\mathrm{FDR}} \ll 0.001$).
  \item According to the barrage rate, cluster~1 had the highest median and cluster~5 the lowest. Pairwise gaps were large (e.g., 1 $>$ 0/5/2/3/6/7, $\Delta$ = 0.50--0.83, all $p_{\mathrm{FDR}} \le 10^{-4}$; and 4 $>$ 0, $\Delta$ = +0.51, $p_{\mathrm{FDR}} \approx 1.2\times10^{-6}$).
  \item According to the like rate, a similar gradient emerged. Specifically, cluster 7 had the highest median, whereas cluster 6 had the lowest. For example, 7 $>$ 6, $+2.70\%$, $p_{\mathrm{FDR}} = 2.26 \times 10^{-6}$.
\end{itemize}

Meanwhile, the video clusters reveal clear differences in interaction characteristics, text features, and video categories. 
In terms of interaction characteristics, cluster 3 represents the most popular group, characterized by the highest number of views. 
This reflects strong user engagement. 
Nevertheless, its normalized interaction rates are not comparatively high, indicating that its popularity is driven primarily by large-scale exposure and video views rather than intensive audience participation.
By contrast, cluster 5 represents a medium-sized group that nonetheless achieves exceptionally high like rates, indicating strong audience stickiness despite its relatively limited reach.
Cluster 7, in turn, is the largest group in terms of video volume and also demonstrates the highest likes and favorite rates, reflecting both a broad audience base and highly intensive engagement.
In addition to clusters 3, 5, and 7, several other groups also display distinctive patterns. 
Cluster 0 demonstrates relatively high likes and favorite rates despite its moderate scale, reflecting stable audience stickiness. 
Cluster 1 is notable for its strong barrage engagement, showing the highest barrage rate across all clusters. 
By contrast, cluster 6 exhibits consistently low interaction rates, marking it as a low engagement group.

In terms of message length, sentiment, and toxicity, each video group also presents different features.
Cluster 6, despite low user interaction, shows the longest message, indicating a small but highly verbose user base. 
Clusters 3 and 4 attract users with relatively high comment activity but also higher toxicity numbers, suggesting that popularity may coincide with contentious discussions. 
Cluster 7, while the most interactive at the video level, is characterized by shorter comments and the most positive sentiment score, reflecting broad but less elaborated participation. 
By contrast, cluster 5 is characterized by medium-scale exposure, relatively short messages, and a high like rate, suggesting a niche but highly appreciative video group. 

\begin{figure}[tp]
\centering
\includegraphics[width=\columnwidth]{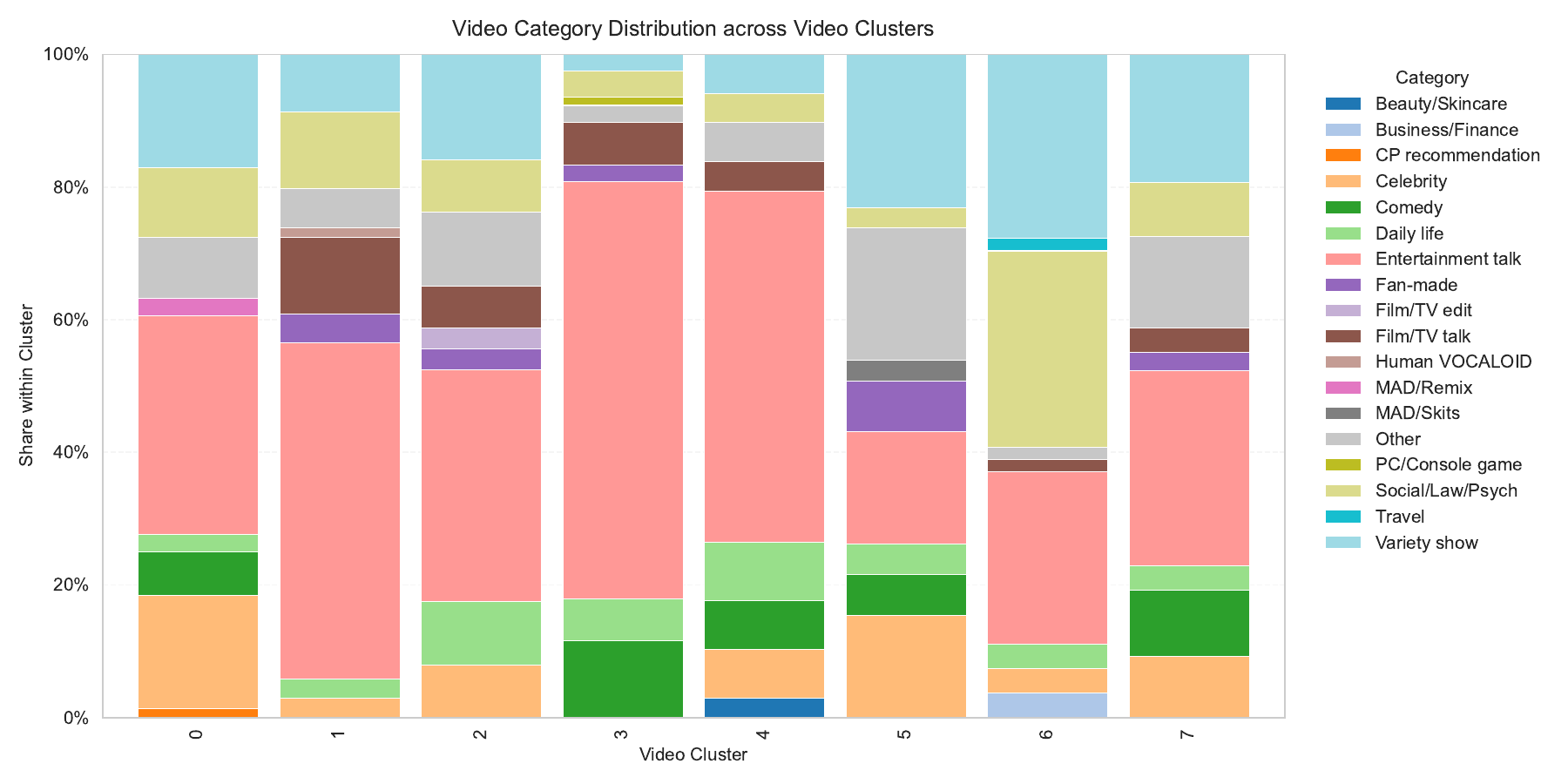}
\caption{Video category distribution across video clusters. 
Bar plots show the distribution of video categories within each video cluster. 
Entertainment categories dominate, while specific clusters (e.g., cluster 6) show higher proportions of Social/Law/Psych and PC/Console games, reflecting thematic differentiation.}
\label{fig:video_category}
\end{figure}

In terms of video categories, the clusters exhibit distinct thematic orientations as shown in Fig.~\ref{fig:video_category}. 
Cluster 3 is highly concentrated in entertainment talk, standing out as the most homogeneous group. 
Cluster 6 differs from the entertainment-dominated groups by featuring a high share of social science and psychology content, alongside variety and finance.
In contrast, cluster 7 presents the most diverse composition, combining entertainment and variety with a wide range of additional genres.

Overall, the video clusters can be grouped into three types. 
The large-scale exposure group, cluster 3, records the highest number of views, the highest level of toxicity, and the most homogeneous video category.
The high-stickiness groups, clusters  7, achieve outstanding engagement, with the highest video number, likes, favorites, and positive sentiment. 
The knowledge-oriented group, cluster 6, records the longest messages and a strong presence of social science content, indicating a small but highly articulate user base.

\subsection{User Clustering Results}

Turning to user clustering, analysis based on the user-by-video interaction matrix identifies nine clusters. 
The clustering evaluation metrics are also summarized in Table~\ref{tab:cluster_quality}.
The silhouette coefficient was 0.353, the Calinski–Harabasz index was 16202.32, and the Davies–Bouldin index was 0.831. 
These results indicate differences among the 9 clusters in terms of interaction scale, sentiment, and toxicity levels.
In addition, Table~\ref{tab:user_cluster_profile} reports user clusters' characteristics based on comment and barrage. 
In this table, ratios are calculated as the proportion of comments or barrage within the total messages.
Then, this study also uses the Kruskal–Wallis test to evaluate each attribute above. 
The test results show that all attributes are significant within each user cluster, as shown in Table~\ref{tab:overall_user}.
The length and comment ratio showed the largest effect sizes. 
And, the  Dunn’s pairwise post-hoc tests with Benjamini–Hochberg FDR correction indicate distinct gradients for both the comment ratio and message length. 
\begin{itemize}
  \item According to the comment ratio. Highest medians in cluster 7, and lowest in 4/5/0. For instance, 7 $>$ 4/0/5 ($\Delta \approx 0.27$, all $p_{\mathrm{FDR}} \ll 0.001$);
  \item According to the message length. Cluster 3 longest, cluster 1/7 is relatively long, and cluster 0 shortest. For example, 3 $>$ 0 ($\Delta = 8.0$, $p_{\mathrm{FDR}} \approx 2.0 \times 10^{-119}$); 1 $>$ 0 ($\Delta \approx 6.77$, $p_{\mathrm{FDR}} \approx 5.9 \times 10^{-104}$).
\end{itemize}
By contrast, sentiment and toxicity showed small effects, suggesting that user-side grouping is driven more by interaction form and text length.

Meanwhile, the user clusters reveal clear differences in message activity, text features, and content preferences. 
In terms of message activity and text features, users in cluster 3 produce the highest number and average length of messages, and the most positive sentiment, suggesting a knowledgeable or rational group. 
Cluster 0 is dominated by very short messages with relatively low sentiment and a high barrage ratio, suggesting a barrage-oriented group focused on rapid interaction. 
Cluster 7, by contrast, shows a relatively high proportion of comments and the lowest toxicity, suggesting a broadly engaged but comparatively friendly group. 
Cluster 8 contains the highest share of comments, while these comments display lower sentiment and moderate toxicity, pointing to a more critical and niche-oriented user base.

\begin{figure}[tp]
\centering
\includegraphics[width=\columnwidth]{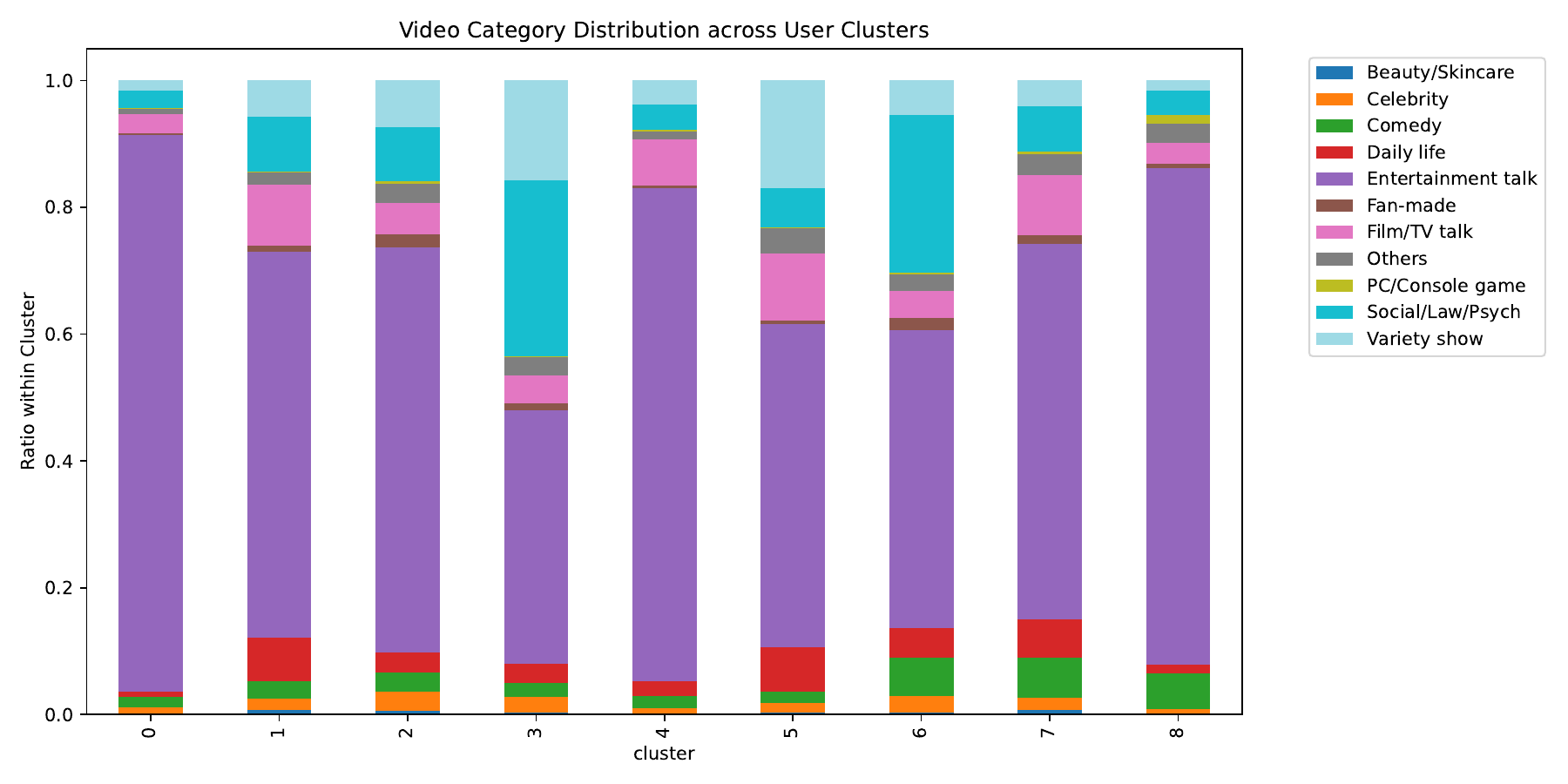}
\caption{Video category distribution across user clusters. 
The stacked bar chart shows the relative share of video categories, the top 10, and Others, within each user cluster. 
Entertainment talk dominates across clusters, but clusters 3 and 6 show higher proportions of Social/Law/Psych or Variety show content, reflecting distinct topical preferences.}
\label{fig:user_cluster_typename}
\end{figure}

In terms of content preferences, Entertainment talk overwhelmingly dominates across all user clusters, though its relative share varies, as shown in Fig.~\ref{fig:user_cluster_typename}. 
Clusters 0, 4, and 8 are the most homogeneous, almost entirely centered on entertainment content. 
Clusters 3, 5, and 6 show more balanced structures, where Social/Law/Psych and Variety categories gain noticeable weight, reflecting mixed interests beyond pure entertainment. 
By contrast, clusters 1, 2, and 7 maintain entertainment as the core theme but integrate additional topics such as daily life, comedy, and film discussion, suggesting more diversified engagement patterns.

Overall, the nine user clusters can be grouped into two representative types.
The mainstream rational participants, clusters 3, 6, 7, are characterized by high activity, long and positive messages, and engagement across both entertainment and social topics.
The reactive entertainment users, clusters 0 and 4, contribute short and emotionally restrained posts, primarily within entertainment-oriented contexts, reflecting fast-paced participation.

\begin{table*}[t]
\caption{User Cluster Profiles Based on Comment and Barrage Behaviors.}
\begin{center}
\begin{tabular}{|c|c|c|c|c|c|c|c|}
\hline
\textbf{Cluster} & \textbf{Users} & \textbf{Total Message} & \textbf{Comment Ratio} & \textbf{Barrage Ratio} & \textbf{Mean Sentiment} & \textbf{Mean Toxicity} & \textbf{Mean Length}  \\
\hline
0  & 2,287 & 21,758 & 0.162 & 0.838 & 0.223 & 0.147 & 16.674 \\
1  & 2,293 & 20,419 & 0.242 & 0.758 & 0.260 & 0.158 & 26.570 \\
2  & 2,107& 21,711 & 0.293 & 0.707 & 0.247 & \textbf{0.160} & 24.267  \\
3  & 1,983 & \textbf{24,280} & 0.206 & 0.794 & \textbf{0.307} & 0.148 & \textbf{27.181} \\
4  & 2,172 & 21,617 & 0.137 & \textbf{0.863} & 0.230 & 0.149 & 19.157 \\
5  & 1,612 & 15,167 & 0.170 & 0.830 & 0.275 & 0.150 & 22.654 \\
6  & 2,241 & 23,566 & 0.221 & 0.779 & 0.277 & 0.157 & 23.486 \\
7  & \textbf{2,472} & 23,832 & 0.317 & 0.683 & 0.288 & 0.141 & 23.658\\
8  & 2,017 & 19,082 & \textbf{0.330} & 0.670 & 0.233 & 0.157 & 22.902\\
\hline
\end{tabular}
\label{tab:user_cluster_profile}
\end{center}
\begin{flushleft}
{\footnotesize \textit{Note:} Ratios are calculated as the proportion of comments or barrage within the total messages.
The values highlighted in bold represent the highest values across each evaluation index.}
\end{flushleft}
\end{table*}

\begin{table}[t]
\caption{Kruskal–Wallis Tests Across User Clusters.}
\begin{center}
\begin{tabular}{|c|c|c|c|}
\hline
\textbf{Variable} & \textbf{H} & \textbf{$p_{\mathrm{overall}}$} & \textbf{$\varepsilon^2$} \\
\hline
Total message   & 332.412 & $5.12\times 10^{-67}$ & 0.017 \\
Comment ratio    & 869.270 & $<10^{-180}$ & 0.045 \\
Sentiment        & 376.960 & $1.58\times 10^{-76}$ & 0.019 \\
Toxicity         & 115.428 & $2.91\times 10^{-21}$ & 0.006 \\
Length       & 876.330 & $<10^{-180}$ & 0.045 \\
\hline
\end{tabular}
\label{tab:overall_user}
\end{center}
\begin{flushleft}
{\footnotesize \textit{Note:} Kruskal-Wallis Tests were conducted at the user level. 
All differences across user clusters are statistically significant. 
}
\end{flushleft}
\end{table}

\section{Discussion}
\subsection{Interpretation of Findings}
This study, based on bidirectional cluster analysis of video and user interaction matrices, identified structural differences in how emotion and toxicity are expressed.
According to the analysis results, both video and user-side clusterings exhibit acceptable quality. 
The Silhouette coefficients (0.348, 0.353) indicate moderate separation with some boundary overlap.
The Calinski–Harabasz indices (692.62, 16,202.32) are high, consistent with compact, well-separated clusters. 
The Davies–Bouldin indices (0.882, 0.831) below 1 further support robustness. 
Thus, these metrics meet common acceptability thresholds while leaving room for sharper boundaries.

Subsequent analysis revealed that interaction scale and comment characteristics at both the video and user levels were systematically associated with where negative interactions tended to appear. 
At the video level, large-scale exposure cluster 3 gained the widest exposure but showed low sentiment and high toxicity.
This supports earlier findings that popular topics with moral or emotional language spread faster and trigger stronger reactions~\cite{brady2017emotion}.
Cluster 3 is also the most homogeneous group. 
In such groups, repeated exposure and algorithmic curation may further amplify controversial views and strengthen echo chambers~\cite{cinelli2021echo}.
Thus, videos with large audiences not only attract attention but also attract more emotionally charged and occasionally hostile expressions.
In contrast, high-stickiness cluster 7 achieved the highest video number, high engagement rates, and maintained a positive atmosphere with low toxicity. 
Such groups show that frequent and stable interaction can create trust and positive expression.
Meanwhile, the knowledge-oriented cluster 6 was small but used long and well-organized messages. 
These users expressed ideas through careful reasoning rather than emotion.
Together, these clusters show that visibility, stability, and expression style are associated with the emotional tone of social video spaces.

At the user level, comment behaviors and sentiment features also differ substantially across groups.
Barrage-oriented clusters rely on short texts, with clusters 0 and 4 marked by rapid interactions, both showing relatively low toxicity.
Comment-oriented clusters emphasize longer texts: cluster 3 produces rational and positive expressions, while cluster 7 combines a broad user base with the lowest toxicity.
These results indicate that text length, sentiment orientation, and comment source are key factors distinguishing user groups.
The above patterns also suggest that different interaction forms shape how users express emotions and communicate with others.

Building on these observations, the findings extend existing research on recommendation exposure.
Prior work on Facebook shows that recommendation algorithms reduce exposure to diverse content, reinforcing information homogenization, while users’ own choices further restrict diversity~\cite{bakshy2015exposure}.
This mechanism has been linked to emotional polarization.
In contrast, the clustering patterns observed in this study reveal how such mechanisms materialize within specific platform interaction structures, particularly in hybrid social–video systems like Bilibili. 

In conclusion, toxic interactions tend to emerge where high exposure co-occurs with homogeneous audiences and emotionally charged themes. 
By contrast, knowledge-oriented and high-interaction/low-toxicity communities are associated with more constructive discourse. 
Thus, exposure scale and audience composition are the primary separators of interaction quality.

\subsection{Governance Implications}
The clustering results indicate that governance should adopt differentiated approaches for different groups.

At the video level, large-scale exposure groups such as cluster 3 require timely intervention during periods of rapid view growth. 
When the inflow of comments reaches a defined threshold, slow mode can be activated, and toxicity detection can be used to limit messages. 
In contrast, high-stickiness groups such as clusters 5 and 7 display stable and positive engagement. 
For these groups, positive incentive measures such as contributor badges and the promotion of co-created content are more appropriate to strengthen community ties.

At the user level, for mainstream rational participants, clusters 3, 6, and 7, platforms should strengthen mechanisms that sustain rational dialogue and encourage cross-topic engagement.
For reactive entertainment users, clusters 0 and 4, interventions should focus on moderating impulsive interactions through gentle feedback and context-rich recommendations.

\subsection{Limitations and Future Directions}
Nonetheless, several limitations remain. 

First, the study relies on data from a single platform. 
While Bilibili is representative, platforms such as YouTube and Twitter differ in user demographics, interaction mechanisms, and governance policies. 
Additionally, platform moderation on Bilibili can significantly reduce extremely toxic content.
Therefore, the cross-platform applicability of our findings requires further verification. 
Future research should expand to multiple platforms and multilingual settings to compare toxic interactions in varied contexts.

Second, there are limitations in sentiment and toxicity indicators.
Toxicity analysis relies on the Perspective API and only uses overall toxicity as the main metric, without distinguishing dimensions such as severe toxicity or threats. 
Meanwhile, sentiment analysis relies on HanLP’s single sentiment score, without capturing multi-dimensional emotions such as anger or joy. 
This simplification may obscure nuanced links between toxicity and emotion.
Future studies could incorporate multi-emotion recognition and multimodal data to enrich analysis.

Finally, the study does not capture dynamic or causal processes.
Our analysis is static, overlooking the evolution of toxic interactions.
On social platforms, toxicity often shows periodic bursts and fluctuations, shaped by recommendation feedback and echo chamber effects.
Future work could integrate time series models and causal inference methods to examine the mechanisms driving group evolution.

\section{Conclusion}
This study investigated the structural and linguistic dynamics of user interaction on social video platforms by modeling video and user relationships as an interaction matrix.
Through separate clustering on both sides of the video and user interaction, the analysis integrated behavioral, text, and context dimensions, allowing for an understanding of how toxicity and sentiment are distributed within platform structures.

The results highlight a clear differentiation between video and user groups.
On the video side, high exposure clusters show intensified emotional polarization and higher toxicity, suggesting that popularity amplifies the risk of harmful discourse.
Conversely, user clusters characterized by longer messages and higher comment ratios demonstrate more rational, constructive participation and lower toxicity.
These patterns suggest that interaction style and visibility jointly shape the emotional climate of online communication.

Building on these findings, the study offers implications for platform governance and the design of recommendation systems.
Platforms should develop early warning mechanisms to monitor rapidly expanding high-visibility videos, where emotional escalation and toxic diffusion are most likely to occur.
Meanwhile, recommendation algorithms should consider and sustain engagement from rational user clusters to strengthen positive discourse and mitigate polarization.
Such differentiated strategies can help maintain a healthier communicative environment without undermining user expression.

However, this study is limited to Bilibili, which restricts its cross-platform generalizability. 
Except for this, this study relies on general sentiment and toxicity indicators. 
The Perspective API measures overall toxicity without distinguishing fine-grained categories such as insults or threats, and HanLP provides a single sentiment score, obscuring complex emotional nuances.

In conclusion, this study advances understanding of toxic interactions on social video platforms.
By analyzing users and videos within a clustering framework, this study connects behavioral and linguistic perspectives, reveals diverse interaction patterns, and offers guidance for promoting constructive and reducing harmful engagement.

\bibliographystyle{IEEEtran}
\bibliography{reference}

\end{document}